\documentclass[twocolumn,showpacs,amsmath,amssymb,prb,superscriptaddress]{revtex4}

\usepackage{graphicx}
\usepackage{bm}
\begin{document}

\title{High-temperature surface superconductivity in topological flat-band systems}

\author{ N.B.\ Kopnin } \affiliation{ Low
Temperature Laboratory, Aalto University, P.O. Box 15100, 00076
Aalto, Finland} \affiliation{ L.~D.~Landau Institute for
Theoretical Physics, 117940 Moscow, Russia}
\author{T.T. Heikkil{\"a}}
\affiliation{ Low Temperature Laboratory, Aalto University, P.O.
Box 15100, 00076 Aalto, Finland}

\author{ G.E. Volovik } \affiliation{ Low
Temperature Laboratory, Aalto University, P.O. Box 15100, 00076
Aalto, Finland} \affiliation{ L.~D.~Landau Institute for
Theoretical Physics, 117940 Moscow, Russia}

\date{\today}

\begin{abstract}
We show that the topologically protected flat band emerging on a
surface of a nodal fermionic system promotes the surface
superconductivity due to an infinitely large density of states
associated with the flat band. The critical temperature depends
linearly on the pairing interaction and can be thus considerably
higher than the exponentially small bulk critical temperature. We
discuss an example of surface superconductivity in multilayered
graphene with rhombohedral stacking.
\end{abstract}

\maketitle

Normal Fermi liquid is the generic form of a system of interacting
fermions. Fermi liquid has a finite density of states (DOS) at
zero energy, which may lead to instabilities at low $T$ with
formation of broken symmetry states with smaller DOS. However,
there is a class of fermionic systems with diverging DOS: the
systems with a dispersionless spectrum that has exactly zero
energy, i.e., the flat band. Historically  this was first
discussed in the context of Landau levels. However, flat bands may
emerge also without a magnetic field, for example in strongly
interacting condensed matter systems
\cite{Khodel1990,NewClass,Shaginyan2010,Gulacsi2010}, in layered
systems with integer-valued pseudospin \cite{Dora2011}, in 2+1
dimensional quantum field theory dual to a gravitational theory in
the anti-de Sitter background \cite{Sung-SikLee2009}, etc.
In some cases the flat band is protected by topology in the
momentum space: topologically protected zero modes emerge in cores
of quantized vortices
\cite{KopninSalomaa1991,Volovik2011,HeikkilaKopninVolovik10}, on
surfaces of gapless topological media such as nodal
superconductors\cite{Ryu2002,SchnyderRyu2010,HeikkilaKopninVolovik10}
and multilayered
graphene\cite{GuineaCNPeres06,MakShanHeinz2010,HeikkilaVolovik10-1,HeikkilaKopninVolovik10},
as well as at the edges of graphene sheets
\cite{Ryu2002,HeikkilaKopninVolovik10}.

In this report we consider a three dimensional (3D) system where
the topologically protected flat band with its singular DOS
appears on the surface giving rise to the 2D surface
superconductivity. This property is generic and does not depend
much on the details of the system. For illustration we use the
multilayered graphene with rhombohedral stacking, where a surface
flat band appears in the limit of large number of layers. We show
that the superconducting critical temperature depends linearly on
the pairing interaction strength and can be thus considerably
higher than the usual exponentially small critical temperature in
the bulk. This may open a new route to room-temperature
superconductivity \cite{HeikkilaKopninVolovik10}. Formation of
surface superconductivity is enhanced already for a system having
$N\geq 3$ layers where the normal-state spectrum has a power-law
dispersion $\xi_{p} \propto |{\bf p}|^N$ as a function of the
in-plane momentum ${\bf p}$. The DOS $\nu(\xi_{ p}) \propto
\xi_{p}^{(2-N)/N}$ has a singularity at zero energy which results
in a drastic enhancement of the critical temperature. We also
demonstrate that doping leads to a suppression of the surface
critical temperature, contrary to its effect on the bulk
superconductivity where the critical temperature is increased
\cite{CastroNeto05,KopninSonin08}.

\paragraph{The model.}

We consider multilayered graphene structure of $N$ layers in the
discrete representation with respect to the interlayer coupling.
We choose the rhombohedral stacking configuration considered in
\cite{GuineaCNPeres06,MakShanHeinz2010,HeikkilaVolovik10-1,HeikkilaKopninVolovik10}
and assume for simplicity that the most important are jumps
between the atoms belonging to different sublattices parameterized
by a single hopping energy $t$. More general form of the
multilayered Hamiltonian can be found in
Refs.~\cite{McClure69,CastroNeto-review}. In the superconducting
case the Hamiltonian has the form of a matrix in the Nambu space.
The Bogoliubov--de Gennes (BdG) equations are
\[
\sum_{j=1}^N \left( \begin{array}{cc} \hat H_{ij}-\mu_i
\delta_{ij} &
\Delta_i \delta_{ij} \\
\Delta_i^*\delta _{ij} & -\hat H_{ij}+\mu_i \delta_{ij}
\end{array}\right)\left(
\begin{array}{c} \hat u_j \\ \hat v_j\end{array}\right)=E \left(
\begin{array}{c} \hat u_i \\ \hat v_i\end{array}\right),
\]
where the sum runs over the layers. The normal-state Hamiltonian
\cite{HeikkilaVolovik10-1}
\begin{equation}
\hat H_{ij}=v_F(\hat {\bm \sigma}\cdot {\bf p})\delta_{i,j}-t\hat
\sigma_+ \delta_{i,j+1} -t\hat \sigma_- \delta_{i,j-1}\ ,
\label{H-norm}
\end{equation}
$\hat{\bm \sigma}=(\hat\sigma_x,\ \hat\sigma_y)$, $\hat \sigma_\pm
= (\hat \sigma_x \pm i\hat \sigma_y)/2$, and $\hat u_i,\ \hat v_i$
are matrices and spinors in the pseudo-spin space associated with
two sublattices. This Hamiltonian acts on the envelope function of
the in-plane momentum ${\bf p}$ taken near one of the Dirac
points, i.e., for $|{\bf p}|\ll \hbar /a$ where $a$ is the
interatomic distance within a layer; $v_F =3 t_0 a/2\hbar $ where
$t_0$ is the the hopping energy between nearest-neighbor atoms
belonging to different sublattices on a layer. The particle-like,
$\hat u_i$, and hole-like, $\hat v_i$, wave functions near the
Dirac point are coupled via the superconducting order parameter
$\Delta_i$ that can appear in the presence of a pairing
interaction. Here we do not specify the nature of pairing which
can be either due to the electron-phonon interaction or due to
other interactions that have been suggested as a source for
intrinsic superconductivity in graphene, see Refs.~\cite{pairing}.
As a reasonable starting point we assume $s$-wave symmetry of the
order parameter and neglect fluctuations for simplicity, though
they could, in principle, be relevant for 2D superconductivity.
The excitation energy for particles and
holes is measured upwards or downwards, respectively, from the
Fermi level which can be shifted with respect to the Dirac point.
We assume that the shifts at the outermost layers may be different
from the bulk chemical potential due to the presence of a surface
charge, i.e., $\mu_i =\mu$ for $i \ne 0,N$ while
$\mu_{1,N}=\mu+\delta\mu_{1,N}$. The order parameter and the Fermi
level shifts $\mu_i$ are scalars in the pseudo-spin space. We
assume that $\Delta_i$ and $\mu_i$ are much smaller than the
inter-layer coupling energy $t>0$, which in turn is $t\ll t_0$.
Usually, $t\sim 0.1\, t_0$ where $t_0\sim 3$ eV
\cite{CastroNeto-review}.

\paragraph{Spectrum.} We decompose the wave function
\begin{equation}
\left(\begin{array}{c} \hat u_n\\ \hat v_n
\end{array}\right)=\left[ \left(\begin{array}{c} \alpha_n^+
\\ \beta_n^+ \end{array}\right) \otimes \hat \Psi^+  +
\left(\begin{array}{c} \alpha_n^-
\\ \beta_n^- \end{array}\right) \otimes \hat \Psi^- \right]\label{wavefunc-u}
\end{equation}
into the spinor functions localized at each sublattice
\[
\hat \Psi_j^+   =\left(\begin{array}{c} 1\\0\end{array}\right)
 \ , \; \hat \Psi_j^-   =\left(\begin{array}{c}
0\\1\end{array}\right)  \ .
\]
The BdG equations take the form
\begin{eqnarray}
\check \tau_3\left[ v_F(p_x-ip_y)\check \alpha_n^- -t\check
\alpha_{n-1}^- -\mu \check \alpha_n^+\right]=E \check \alpha_n^+
 ,\; n\ne 1, \quad \label{BdGeqD}\\
\check \tau_3 \left[ v_F(p_x+ip_y)\check \alpha_n^+ -t\check
\alpha_{n+1}^+ -\mu \alpha_n^- \right]=E \check \alpha_n^-  ,\;
n\ne N. \quad \label{BdGeqA}
\end{eqnarray}
We introduce matrices and vectors in the Nambu space
\[
\check \tau_3=\left(\begin{array}{lr} 1 & 0 \\ 0&
-1\end{array}\right)\ , \; \check \Delta_n
=\left(\begin{array}{lr} 0 & \Delta_n \\ \Delta^*_n &
0\end{array}\right)\ , \; \check \alpha_n^\pm =
\left(\begin{array}{c} \alpha_n^\pm
\\ \beta_n^\pm  \end{array}\right)\ .
\]
In Eqs.~(\ref{BdGeqD}) and (\ref{BdGeqA}) we assume that $\Delta_n
\ne 0$ only at the outermost layers, while $\Delta_n =0$ for $n\ne
1,N$. The arguments supporting this model are given below. We also
neglect $\Delta_n$ as compared to $t$ in Eqs.~(\ref{BdGeqD}) and
(\ref{BdGeqA}) for $n=N$ and $n=1$, respectively, as they lead to
higher-order corrections in $\Delta/t$. The particle and hole
channels are thus decoupled if $n\ne 1,N$ which determines the
coefficients $\check \alpha_n^{\pm} =\check A^\pm e^{ip_z d n}$
and the energy in terms of the transverse momentum $p_z$ ($d$ is
the interlayer distance) \cite{HeikkilaVolovik10-1}
\begin{equation}
E^2=v_F^2 p^2 -2tv_Fp \cos(p_z d -\phi)+t^2 \label{E-bulk}
\end{equation}
where $p=\sqrt{p_x^2+p_y^2}$ and $e^{i\phi}=(p_x+ip_y)/p$.

A finite order parameter $\Delta$ couples the particle and hole
channels at the outermost layers, $i=1$ and $i=N$,
\begin{eqnarray}
\check \tau_3 v_F(p_x-ip_y)\check \alpha_1^- -\check \tau_3
\mu_1\check \alpha_1^+ &=&E \check \alpha_1^+
-\check \Delta _1 \check \alpha_1^+ \label{eqn-u=1}\ , \\
\check \tau_3 v_F(p_x+ip_y)\check \alpha_N^+ -\check \tau_3
\mu_N\check \alpha_N^- &=&E \check \alpha_N^- -\check \Delta _N
\check \alpha_N^- \ . \label{eqn-u=N}
\end{eqnarray}

Boundary conditions (\ref{eqn-u=1}), (\ref{eqn-u=N}) select $p_z$
and determine $2N$ particle and hole branches of the energy
spectrum. Looking for the branches that belong to the surface
states with energies of the order of $\Delta$ and $\mu$, we solve
these equations for $E\ll t$. Since Eqs.~(\ref{BdGeqD}),
(\ref{BdGeqA}) do not contain $\Delta$, one can use the
coefficients as obtained in Ref. \cite{HeikkilaVolovik10-1}
\begin{eqnarray*}
\check \alpha^+_n &=&\frac{C}{\sqrt{2}}\left[ \left(
\frac{v_Fp}{t}\right)^{n-1}\check A^+ \right. \\
&&+\left.\left( \frac{v_Fp}{t}\right)^{N-n}\frac{v_Fp(\check
\tau_3 E+\mu)
}{v_F^2p^2-t^2} \check A^-\right]e^{i(n-1-\frac{N}{2})\phi}\ , \\
\check \alpha^-_n &=&\frac{C}{\sqrt{2}}\left[\left(
\frac{v_Fp}{t}\right)^{N-n} \check A^- \right. \\
&&+\left. \left( \frac{v_Fp}{t}\right)^{n-1}\frac{v_Fp(\check
\tau_3 E+\mu) }{v_F^2p^2-t^2}\check
A^+\right]e^{i(n-\frac{N}{2})\phi}\ .
\end{eqnarray*}
Here $C$ is a normalization constant. We include the first-order
corrections in energy. Having an imaginary momentum $p_z$ for
$v_Fp<t$, these solutions decay away from the surfaces  and thus
they describe the surface states. The vectors $ \check A^\pm =
\left(A^\pm , \, B^\pm \right)^T$ do not depend on $n$. Equations
(\ref{eqn-u=1}) and (\ref{eqn-u=N}) yield
\begin{eqnarray}
\check \tau_3 \xi_{ p} \check A^-= (\tilde E+\check \tau_3 \tilde
\mu_1)\check A^+ - \check
\Delta_1 \check A^+ \ , \label{BdGeq-mu1}\\
\check \tau_3 \xi_{ p} \check A^+= (\tilde E+\check \tau_3 \tilde
\mu_N)\check A^- - \check \Delta_N \check A^- \ ,
\label{BdGeq-mu4}
\end{eqnarray}
where $\xi_{p} =t \left( v_Fp/t\right)^N$, $\tilde
\mu_{1,N}=\tilde \mu +\delta\mu_{1,N}$, and
\begin{equation}
 E=\tilde E(1-v_F^2p^2/t^2) , \; \mu =\tilde \mu(1-v_F^2p^2/t^2)\
 .
\label{tildeE}
\end{equation}
Equations (\ref{BdGeq-mu1}), (\ref{BdGeq-mu4}) provide the
surface-state spectrum
\begin{eqnarray}
\left[ \tilde E^2-\tilde \mu_N^2-|\Delta_N|^2\right]
\left[\tilde E^2-\tilde\mu_1^2-|\Delta_1|^2\right]+\xi_{ p}^4 \nonumber \\
-\xi_{ p}^2\left[2\tilde E^2 +2\tilde \mu_1\tilde \mu_N
-\Delta_1^*\Delta_N - \Delta_1\Delta_N^*\right]=0 \ .
\label{cond1}
\end{eqnarray}

\begin{figure}[b] \centering
\includegraphics[width=\columnwidth]{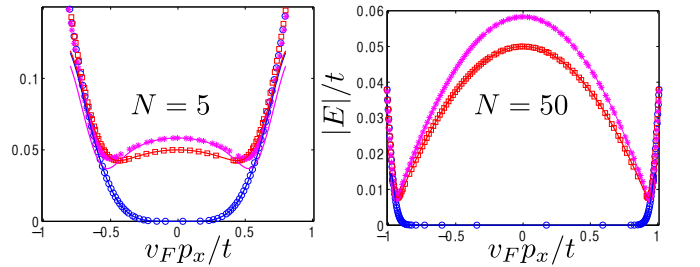}
\caption{(Color online): Spectrum of surface states for different
numbers of layers. Left: $N=5$ and right: $N=50$. The symbols have
been calculated by exact diagonalization and solid lines are
computed from Eq.~\eqref{E-mu} up to the point where the
approximation in it is valid. The three cases are: normal case
with $\mu_n \equiv 0$ (blue circles), $\Delta=0.05 t, \mu_n \equiv
0$ (red squares), and $\Delta=0.05 t, \mu_1=\mu_N=0.03 t$ (magenta
crosses).} \label{fig:spectrum}
\end{figure}

If $\Delta_1=\Delta_N$ we have from Eq.~(\ref{cond1})
\begin{equation}
\tilde E^2_\pm =\left[ \tilde \mu_0\mp \sqrt{ \xi_{ p}^2
+(\delta\mu)^2} \right]^2 +|\Delta|^2\ ,  \label{E-mu}
\end{equation}
where $\delta \mu =(\mu_1-\mu_N)/2$ and $\tilde \mu_0=(\tilde
\mu_1+\tilde \mu_N)/2$. Equations (\ref{BdGeq-mu1}),
(\ref{BdGeq-mu4}) determine four independent states. If $\mu
=\delta\mu=0$ they are (i) $\tilde E_1=\tilde E$ and $A_1^\pm = u
$, $B_1^\pm = v $, (ii) $\tilde E_2=- \tilde E$ and $A_2^\pm = v
$, $B_2^\pm =- u $, (iii) $\tilde E_3=\tilde E_0$ and $A_3^\pm
=\pm   v $, $B_3^\pm =\pm  u$, (iv) $\tilde E_4=-\tilde E$ and
$A_4^\pm =\pm u $, $B_1^\pm =\mp v $. Here $\tilde E=
\sqrt{\xi_p^2 +\Delta^2}$ and
\begin{equation}
u=\frac{1}{\sqrt{2}}\left[ 1+ \xi_{ p}/\tilde E
\right]^\frac{1}{2} , \; v=\frac{1}{\sqrt{2}}\left[ 1- \xi_{
p}/\tilde E \right]^\frac{1}{2}\ .\; \label{A}
\end{equation}
The overall normalization requires $d \sum_{n=1}^N[|\alpha^+_n|^2
+|\beta^+_n|^2+|\alpha^-_n|^2 +|\beta^-_n|^2]=1$. For $\xi_{ p}\ll
t$ this gives
\begin{equation}
|C |^2=d^{-1}[1-(v_Fp/t)^2]\ . \label{norm}
\end{equation}
Note that Eqs. (\ref{cond1})--(\ref{norm}) hold for $\xi_{ p}\ll
t$. The spectrum is plotted in Fig.~\ref{fig:spectrum}.

If $N\to \infty$ and $\xi_{p}\to 0$ for any $v_F p/t < 1$, the
surface-state part localized at $n=N$ (with the coefficients
$\check A^-$) decouples from that (with $\check A^+$) which is
localized at $n=1$. For a ``flat band'' $\xi_{ p} \to 0$,
Eq.~(\ref{cond1}) yields
\begin{equation}
\tilde E^2_+=\tilde \mu_N^2 +|\Delta_N|^2 ~{\rm or}~ \tilde
E^2_-=\tilde \mu_1^2 +|\Delta_1|^2 \ . \label{E-mu-flb}
\end{equation}
This shows that the definite signs in Eq.~(\ref{E-mu}) belong to
the surface states localized at the corresponding layers.


\paragraph{Flat band; zero doping.}

The gap at a layer $N$ is
\begin{eqnarray*}
\Delta_N&=& \int \frac{d^2 p}{(2\pi \hbar)^2} \sum_{k=1}^N
V_{{\bf p},p_z(k)}{\rm Tr}\, [\hat u_N({\bf p},k) \hat v^*_N({\bf p},k)]\\
&&\times [1-2f(E_{{\bf p}, k})], \label{OP1}
\end{eqnarray*}
where $f(E)$ is the Fermi distribution function. We assume that
the cut-off momentum $p_c$ of the pairing potential $V$ is larger
than $p_{\rm FB}=t/v_F$. The sum includes one $n=N$ surface state
which we label by $k=0$ and the bulk states specified by the
transverse momenta $ p_z(k)$, where $k=1,2,\ldots, N-1 $ with the
spectrum of Eq.~(\ref{E-bulk}). Therefore, $\Delta_{N}=\Delta_S
+\Delta_B$ where the surface contribution comes from the flat band
area $p<p_{\rm FB}$,
\begin{eqnarray}
\Delta_S&=& V \int_{p<p_{\rm FB}} \frac{d^2 p}{(2\pi \hbar)^2}
{\rm Tr}\, [\hat u_N({\bf p},0) \hat v^*_N({\bf p},0)] \nonumber \\
&&\times [1-2f(E_{{\bf p}, 0})]. \ \label{Delta-S}
\end{eqnarray}
The bulk contribution comes from the momenta $p>p_{\rm FB}$. For
such momenta, the surface state $k=0$ will also extend to the bulk
giving rise to (for $T=0$)
\begin{eqnarray}
\Delta_{B}= V \int_{p_{\rm FB}<p<p_c} \frac{d^2 p}{(2\pi \hbar)^2}
\left\{ {\rm Tr}\, [\hat u_N({\bf p},0) \hat v^*_N({\bf p},0)]
\right.
\nonumber \\
+\left. \sum_{k=1}^{N-1} {\rm Tr}\, [\hat u_N({\bf p},k) \hat
v^*_N({\bf p},k)]\right\}\ .
\label{Delta-B}
\end{eqnarray}

All the bulk states with $p>p_{\rm FB}$ are normalized to the
sample width $W=dN$, i.e., $ u^*(z)\sim 1/\sqrt{W}$. According to
Eq.~(\ref{E-bulk}), $E \sim v_F p>t$ in
Eq.~\eqref{Delta-B}.
Therefore,
\[
\Delta_B \approx \frac{Vp_c^2}{4\pi \hbar^2}\frac{N}{W}\left[
\frac{\Delta}{v_F p_c}- {\cal O}\left(\frac{\Delta}{v_F
p_c}\right)^3 \right]\ .
\]
If there was only the bulk contribution ($\Delta \equiv
\Delta_B=\Delta_N$), the gap equation would have a nonzero
solution only for a potential strength higher than a certain
critical value $ Vp_c/4\pi \hbar^2 v_F d  >1 $, as is the case in
the usual single-layer graphene with zero
doping \cite{CastroNeto05,KopninSonin08}.

The surface states for $p<p_{\rm FB}$ are normalized according to
Eq.~(\ref{norm}). We find from Eq.~(\ref{Delta-S})
\begin{equation}
\Delta_S= 2 V \int_{p<p_{\rm FB}} \frac{d^2 p}{(2\pi
\hbar)^2}|C|^2 uv \tanh\frac{E}{2k_B T} \ . \label{self2}
\end{equation}
For simplicity we assume that $ V$ is constant up to the cut-off
momentum $p_c$. Here $u$ and $v$ are determined by Eqs.~(\ref{A}),
(\ref{norm}). In the case of a flat band $uv=1/2$ while $E=\Delta
(1-v_F^2p^2/t^2)$. For $T=0$ it gives
\begin{equation}
\Delta_S=\Delta_0 \equiv g/8\pi  \ , \label{Delta-flat}
\end{equation}
where $g=\tilde V p_{\rm FB}^2/\hbar^2$ is the characteristic
pairing energy, $\tilde V = V/d$ is the two-dimensional pairing
potential.

The ratio of the order parameter in the bulk to that on surface is
of the order $ (\Delta/t)(v_Fp_c/t) $. Since $\Delta \ll t$, the
contribution from the bulk states with $E>t$ can be neglected if
the cut-off momentum of the interaction $p_c$ does not
considerably exceed $t/v_F$. We thus arrive at the central result
of our paper, namely that {\it the surface superconductivity in
the presence of a flat band dominates over the bulk
superconductivity}. This follows from an infinitely large density
of states associated with the flat band. The critical temperature
is determined by Eq.~(\ref{Delta-flat}) with $\Delta \to 0$, which
gives $\Delta_0=3k_B T_c$. Due to its linear dependence on the
interaction strength, {\it the critical temperature is
proportional to the area of the flat band and can be essentially
higher than that in the bulk}.

For a flat band $\xi_{ p} =0$ with $p_c=p_{\rm FB}$ the only
characteristic values in the superconducting surface state are the
energy $\Delta$ and the momentum $p_{\rm FB}$. Therefore, the
coherence length should be of the order of the only available
length scale, $ \xi_0 \sim \hbar/p_{\rm FB} $. It is much larger
than the interatomic distance, $\xi_0 \gg a$, since $p_{\rm FB}
\ll p_0\sim \hbar /a$.

Doping destroys the surface superconductivity. This can be seen
from Eq.~(\ref{self2}) with $uv= \Delta/2 \tilde E_+$ and
$E=(1-v_F^2p^2/t^2)\tilde E_+$ where $\tilde E_+$ is taken from
Eq.~(\ref{E-mu-flb}). The critical temperature is found by putting
$\Delta=0$. For example, if $\mu$ and $\mu_N$ have the same sign,
both $\Delta_0$ and $T_c$ vanish at the critical doping level that
satisfies
\[
1=\frac{\tilde V}{4\pi \hbar^2 |\mu_N
-\mu|}\left|\frac{1}{2}-\frac{\mu}{\mu_N -\mu}+\frac{\mu^2}{(\mu_N
-\mu)^2}\ln\frac{\mu_N}{\mu}\right|\ .
\]
If $\mu_N =\mu$ the critical doping is $|\mu| =2k_B T_c$.

\paragraph{Surface superconductivity in a finite array.}

Since the normal-state DOS defined as
\begin{equation}
\nu(\xi_p) = \frac{p}{2\pi\hbar^2}\,
\frac{dp}{d\xi_p}=\frac{t(\xi_p /t)^{\frac{2-N}{N}}}{2\pi \hbar^2
N v_F^2}\  \label{DOS}
\end{equation}
has a low-energy singularity for $N>2$, the surface
superconductivity is favorable already for a system with a finite
number of layers $N\geq 3$. A simple expression for the
zero-temperature gap can be obtained if $N\geq 5$. For a finite
$N$, the value $\xi_p$ can reach values larger than $\Delta$. We
use Eqs. (\ref{E-mu}) - (\ref{norm}) for zero doping in
Eq.~(\ref{self2}) where the upper limit of integration $p_c$ is
now such that $\xi_c=t(v_Fp_c/t)^N \gg \Delta$. Transforming to
the energy integral with the normal-state DOS Eq.~(\ref{DOS}) we
see that, for $N>4$, the integral converges at $\xi_p \sim \Delta$
or $p\sim p_\Delta =p_{\rm FB}(\Delta/t)^\frac{1}{N}$. The
zero-temperature gap is
\begin{equation}
\Delta_0 =t\left(\frac{ g }{4\pi
t}\left[\alpha(N)-\frac{1}{2}\left(\Delta_0/t\right)^\frac{2}{N}
\alpha(N/2)\right] \right)^\frac{N}{N-2}
\end{equation}
where
\[
\alpha(N) =\int_0^{\infty} \frac{x^\frac{N+2}{N}\,
dx}{\sqrt{x^2+1}^3}=\frac{1}{\sqrt{\pi}}\Gamma\left(\frac{N-2}{2N}\right)
\Gamma\left(\frac{N+1}{N}\right)\ .
\]
For $N\gg 1$ we have $\alpha_N=1$. The flat-band result,
Eq.~(\ref{Delta-flat}), is recovered if the number of layers is
$N\gg 2\ln (t/\Delta_0)$. The coherence length for a finite system
is $\xi_0\sim \hbar /p_\Delta$. It approaches $\hbar/p_{\rm FB}$
for $N\to \infty$.

\begin{figure}[t] \centerline{
\includegraphics[width=0.5\columnwidth]{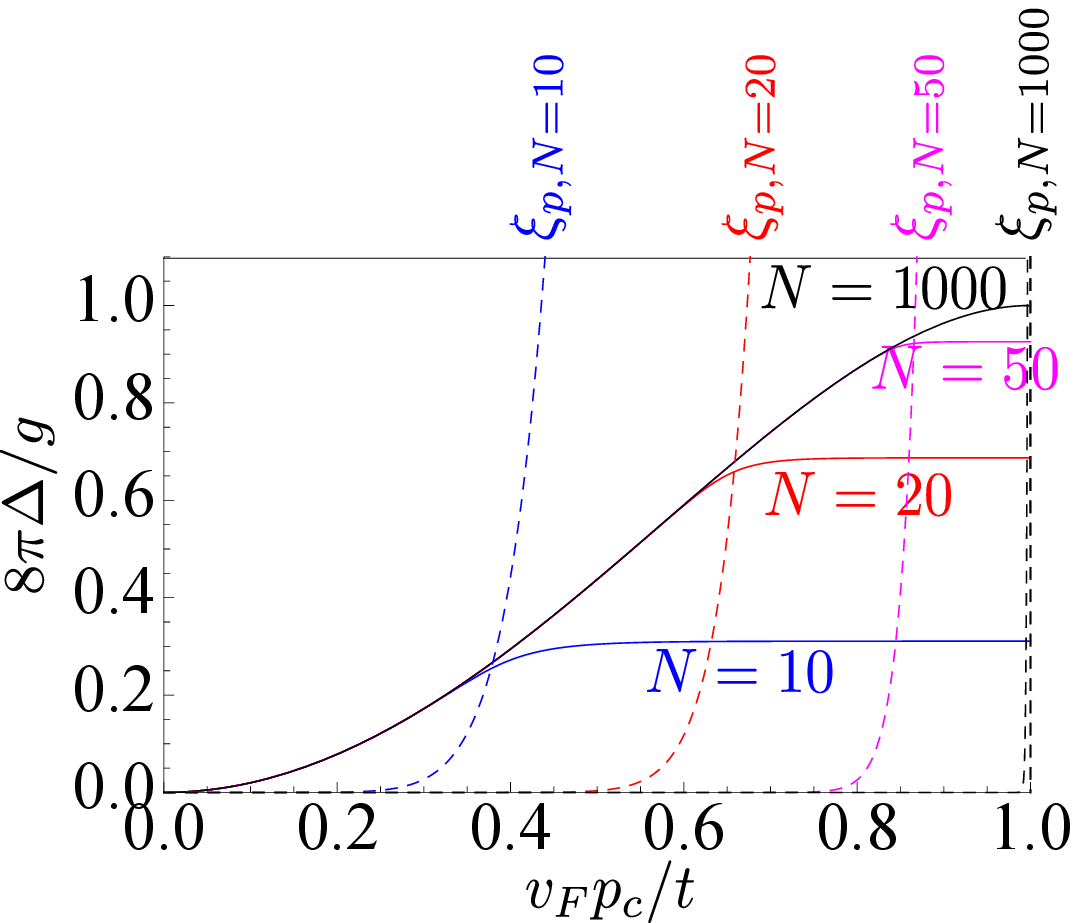}
\includegraphics[width=0.45\columnwidth]{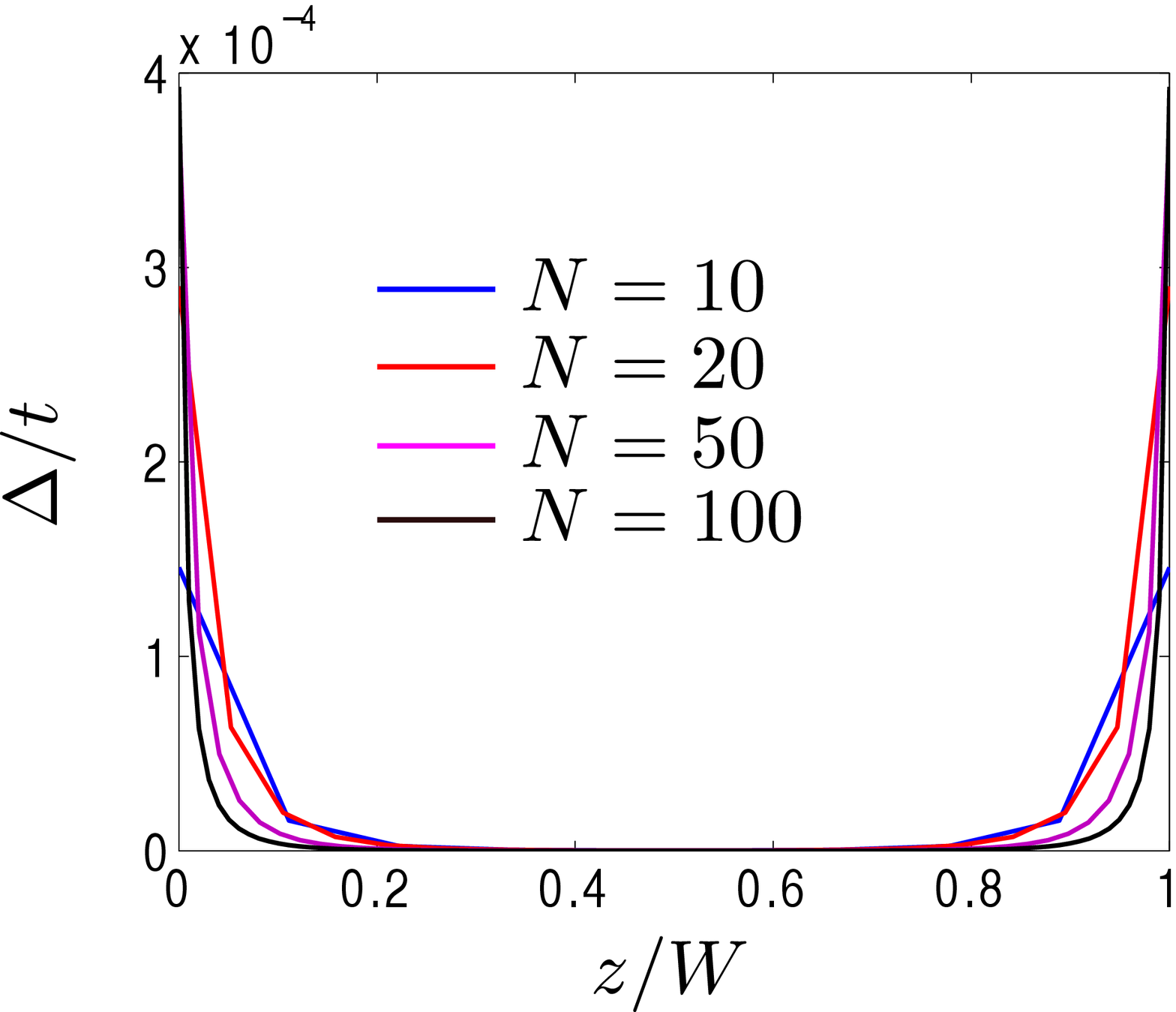}}
\caption{(Color online) Left panel: Zero-temperature gap as a
function of the momentum cutoff $p_c$ for various $N$ (solid
lines). The gap saturates at $p_c\sim p_\Delta$ and approaches
Eq.~(\protect\ref{Delta-flat}) for $N\to \infty$. The dashed lines
show the dispersion $\xi_p$ for each $N$. Right panel: the
self-consistently calculated $\Delta(z)$ profile at different
layers, $z=nd$. On both panels $g=0.01t$.} \label{fig-gap}
\end{figure}

The gap obtained by numerical integration of Eq.~(\ref{self2})
with a cut-off $p_c$ is plotted in Fig.~\ref{fig-gap}, left panel.
The right panel of Fig.~\ref{fig-gap} shows the order parameter as
a function of the transverse coordinate. It extends into the bulk
only over a few interlayer distances due to a decay of the wave
functions. Taking this into account we have chosen the model, Eqs.
(\ref{BdGeqD})--(\ref{eqn-u=N}), in which the order parameter is
nonzero only on the outermost layers.

\paragraph{Conclusion.}

The flat band with infinite DOS emerges in semi-metals with
topologically protected nodal lines. The flat band promotes
surface superconductivity with $T_c$ proportional to the pairing
interaction strength and to the area of the flat band in the
momentum space which is determined by the projection of the nodal
line onto the surface. The critical temperature can thus be
considerably higher than the exponentially small $T_c$ in the
bulk. Formation of surface superconductivity is enhanced already
for a system with a number of layers $N\geq 3$ where the normal
DOS has a singularity at zero energy. Topologically protected flat
bands may also appear on interfaces, twin boundaries and grain
boundaries in bulk 3D topological materials leading to an enhanced
bulk $T_c$. Indications towards surface superconductivity have
been seen in experiments on
graphite\cite{Kopelevich01,Esquinazi08}. The enhanced
superconducting density has been reported on twin boundaries in
Ba(Fe$_{1-x}$Co$_x$)$_2$As$_2$ \cite{Moler2010}. These
observations might be explicable with our theory. Our predictions
may be used for search or for artificial fabrication of layered
and/or twinned systems with high- and even room-temperature
superconductivity.

\acknowledgements

We thank A.\ Geim, V.\ Khodel, and K.\ Moler for helpful comments.
This work is supported in part by the Academy of Finland and its
COE program 2006--2011, by the European Research Council (Grant
No. 240362-Heattronics), by the Russian Foundation for Basic
Research (grant 09-02-00573-a), and by the Program ``Quantum
Physics of Condensed Matter'' of the Russian Academy of Sciences.

\end{document}